\documentclass[final,5p,twocolumn]{elsarticle}

\usepackage{graphicx}          
\usepackage[dvips]{epsfig}    

\usepackage{xcolor}
\usepackage{amssymb}
\usepackage{amsthm}
\usepackage{amsmath}
\usepackage{algorithmic}

\begin{document}

\begin{frontmatter}

\title{Convergent dynamics of optimal nonlinear damping control} 

\author{Michael Ruderman}
\ead{michael.ruderman@uia.no}

\address{University of Agder, 4604-Norway}

\begin{abstract}                          
Following Demidovich's concept and definition of convergent
systems, we analyze the optimal nonlinear damping control,
recently proposed \cite{ruderman2021} for the second-order
systems. Targeting the problem of output regulation,
correspondingly tracking of $\mathcal{C}^1$-trajectories, it is
shown that all solutions of the control system are globally
uniformly asymptotically stable. The existence of the unique limit
solution in the origin of the control error and its time
derivative coordinates are shown in the sense of Demidovich's
convergent dynamics. Explanative numerical examples are also
provided along with analysis.
\end{abstract}


\end{frontmatter}

\newtheorem{thm}{Theorem}
\newtheorem{defn}{Definition}
\newtheorem{clr}{Corollary}
\newdefinition{rmk}{Remark}
\newdefinition{exmp}{Example}
\newdefinition{prop}{Proposition}
\newproof{pf}{Proof}

\section{Introduction}
\label{sec:1}

Unlike the linear feedback systems, the nonlinear controllers and
plants require, mostly, some non-universal methods and tools for
analysis and design. Depending on the complexity of underlying
nonlinear dynamics, the local or global approach, and (not least)
the control specification, e.g. whether a set-point or output
regulation \cite{isidori1995} problem is of interest, the
stability and convergence analysis can require a more specific
approach than that based on the standard Lyapunov stability theory
\cite{khalil2002,sastry2013}. For analyzing the behavior of
nonlinear control systems subject to external inputs, such as
reference or disturbance signals, one needs to prove the existence
and global asymptotic stability of a solution, along which the
output regulation is ensured. The so-called incremental stability
\cite{angeli2002} and contraction analysis \cite{lohmiller1998},
or more generally contraction theory, see e.g. \cite{sontag2010},
provide the means to show that all system trajectories converge
uniformly to a unique solution for which the output regulation
error is zero.

Being motivated by the Demidovich's \cite{demidovich1967} concept
and definition of the convergent systems, reviewed in
\cite{pavlov2004}, this note analyzes and develops the output
regulation properties of the optimal nonlinear damping control
\cite{ruderman2021}.

\section{Preliminaries}
\label{sec:2}

\subsection{Convergent dynamics}
\label{sec:2:sub:1}

In this section, we briefly recall the main definition and
properties of a convergent system dynamics, according to
Demidovich \cite{demidovich1967}, while we will closely follow the
developments and notations provided in \cite{pavlov2004}.

For a large class of $n$-dimensional nonlinear systems
\begin{equation}\label{eq:1}
   \dot{x} = f(x,t),
\end{equation}
where the state vector $x \in \mathbb{R}^n$, with $2 \leq n <
\infty$, is continuous in time $t$, and $f(\cdot)$ is the vector
field which is differentiable in $x$, the following notion of
\emph{convergent systems} can be given according to
\cite{demidovich1967}.

\begin{defn}
\label{def:1}
The system \eqref{eq:1} is said to be convergent if
for all initial conditions $t_0 \in \mathbb{R}$, $\bar{x}_0 \in
\mathbb{R}^n$ there exists a solution $\bar{x}(t) =
x(t,t_0,\bar{x}_0)$ which satisfies:
\begin{enumerate}
    \item[(i)] $\bar{x}(t)$ is well-defined and bounded for all $t \in (-\infty,
    \infty)$;
    \item[(ii)]$\bar{x}(t)$ is globally asymptotically
    stable.
\end{enumerate}
\end{defn}
Such solution $\bar{x}(t)$ is called a \emph{limit solution}, to
which all other solutions of the system \eqref{eq:1} converge as
$t \rightarrow \infty$. In other words, all solutions of a
convergent system 'forget' their initial conditions after some
transient time, which is depending on the exogenous values like
reference or disturbance signals, and converge asymptotically to
$\bar{x}(t)$.

\begin{rmk}
\label{rmk:1} If a globally asymptotically stable limit solution
exists, it may be non-unique. Yet if $\bar{x}(t)$ is the single
solution defined and bounded for all $t \in (-\infty, \infty)$,
then the system \eqref{eq:1} is said to be \emph{uniformly
convergent}.
\end{rmk}

The unform convergence requires the system \eqref{eq:1} to have an
unique limit solution $\bar{x}(t)$, like in case of linear
systems. Otherwise, a non-uniformly convergent system might have
also another globally asymptotically stable solutions
$\tilde{x}(t)$, bounded for all $t \in (-\infty, \infty)$, so that
for any such pair of solutions it is valid $\bigl\| \bar{x}(t) -
\tilde{x}(t) \bigr\| \rightarrow 0 $ as $t \rightarrow \infty$.

\begin{rmk}
\label{rmk:2} The system \eqref{eq:1} is, moreover,
\emph{exponentially convergent} if it is uniformly convergent and
the limit solution $\bar{x}(t)$ is globally exponentially stable.
\end{rmk}

For further details on uniform, asymptotic, and exponential
properties of stability of the system solutions we refer to
seminal literature, e.g. \cite{khalil2002,sastry2013}. The
existence and uniqueness of a limit solution of system
\eqref{eq:1} has an essential application to the output regulation
problems \cite{isidori1995,huang2004}. Here, for a given reference
signal of the closed-loop system, one can seek for demonstrating
the control system \eqref{eq:1} is convergent, i.e. has an
asymptotically stable limit solution along which the regulated
output control error is zero. The property of a system to be
convergent follows from the sufficient condition, given by
Demidovich \cite{demidovich1967}, which is formulated in the
following theorem \cite{pavlov2004}:

\begin{thm}
\label{thm:1} Consider the system \eqref{eq:1}. Suppose, for some
positive definite matrix $P=P^T > 0$ the matrix
\begin{equation}\label{eq:2}
J(x,t) := \frac{1}{2} \Bigl ( P \, \frac{\partial f}{\partial
x}(x,t) + \Bigl[ \frac{\partial f}{\partial x}(x,t) \Bigr]^T P
\Bigr)
\end{equation}
is negative definite uniformly in $(x,t) \in \mathbb{R}^n \times
\mathbb{R}$ and $|f(0,t)| \leq \mathrm{const} < +\infty$ for all
$t \in \mathbb{R}$. Then the system \eqref{eq:1} is convergent.
\end{thm}
The proof of the Theorem \ref{thm:1} can be found in
\cite{pavlov2004}. Note that a particular case $f(0,t) \equiv 0$
of the Theorem \ref{thm:1} implies the well celebrated Krasovskii
stability theorem \cite{krasovskii1963}. The uniform negative
definiteness of \eqref{eq:2} implies a vanishing difference
between any two solutions $x_{i}(t)$ and $x_{ii}(t)$ of the system
\eqref{eq:1}, while for an exponentially convergent system (cf.
Remark \ref{rmk:2}) that means
\begin{equation}\label{eq:3}
\bigl|x_{i}(t) - x_{ii}(t)\bigr| <  \alpha \exp\bigl(-\beta
(t-t_0) \bigr) \bigl|x_{i}(t_0) - x_{ii}(t_0)\bigr|
\end{equation}
for all $t > t_0$, where $\alpha,\beta > 0$ have the same values
for all pairs of the solution $\bigl[x_{i}(t)$, $x_{ii}(t)\bigr]$.

\subsection{Optimal nonlinear damping control}
\label{sec:2:sub:2}

In the following, we will summarize the optimal nonlinear damping
control, insofar as necessary for the main results presented in
Section \ref{sec:3}, while for that recently introduced control
and its properties we refer to \cite{ruderman2021}.
\begin{figure}[!h]
\centering
\includegraphics[width=0.9\columnwidth]{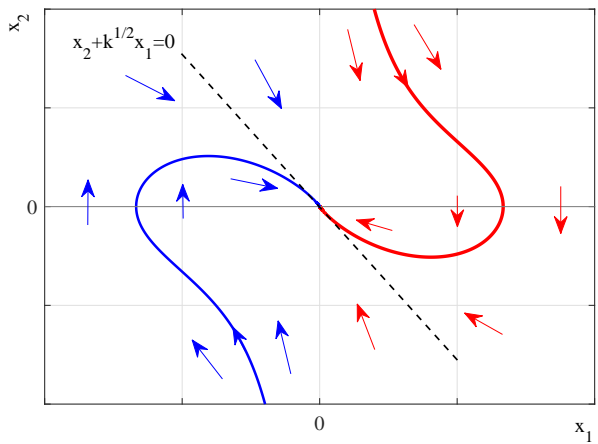}
\caption{Phase portrait of the control system \eqref{eq:4},
\eqref{eq:5}.} \label{fig:1}
\end{figure}

The second-order closed-loop control system with an optimal
nonlinear damping is written as
\begin{eqnarray}
\label{eq:4}
  \dot{x}_1 &=&x_2,  \\
  \dot{x}_2 &=&-kx_1-x_2^2|x_1|^{-1} \mathrm{sign}(x_2), \label{eq:5}
\end{eqnarray}
where $k > 0$ is the single arbitrary design parameter. The system
\eqref{eq:4}, \eqref{eq:5}, where $x_1$ is the output of interest,
is globally asymptotically stable and converges to the unique
equilibrium in the origin: (i) along the attractor
\begin{equation}\label{eq:6}
x_2+ \sqrt{k} x_1=0
\end{equation}
of trajectories in vicinity of the origin, and (ii) without
crossing the $x_2$-axis, see Figure \ref{fig:1}. Note that the
(ii)-nd property prevents singularity (owing to $x_1=0$) in the
solutions of \eqref{eq:4}, \eqref{eq:5}, and that for all initial
conditions $x_1(t_0) \neq 0 \: \wedge \: x_2(t_0) \in \mathbb{R}$
and all trajectories outside the origin i.e. $\|x_1,x_2\|(t) \neq
0$. The control system \eqref{eq:4}, \eqref{eq:5} assumes an
unperturbed system dynamics, so that the robustness against
external perturbations is the subject of future research. Other,
here relevant remarks follow.

\begin{rmk}
\label{rmk:3} The output convergence of the control system
\eqref{eq:4}, \eqref{eq:5} is quadratic on the logarithmic scale
$\log|x_1|$. Hence, the control constitutes a fast alternative to
the standard proportional-derivative controller which, with the
same proportional gain factor $k$, converges only linearly on the
logarithmic scale $\log|x_1|$, cf. \cite[Fig.~5]{ruderman2021}.
\end{rmk}

\begin{figure}[!h]
\centering
\includegraphics[width=0.9\columnwidth]{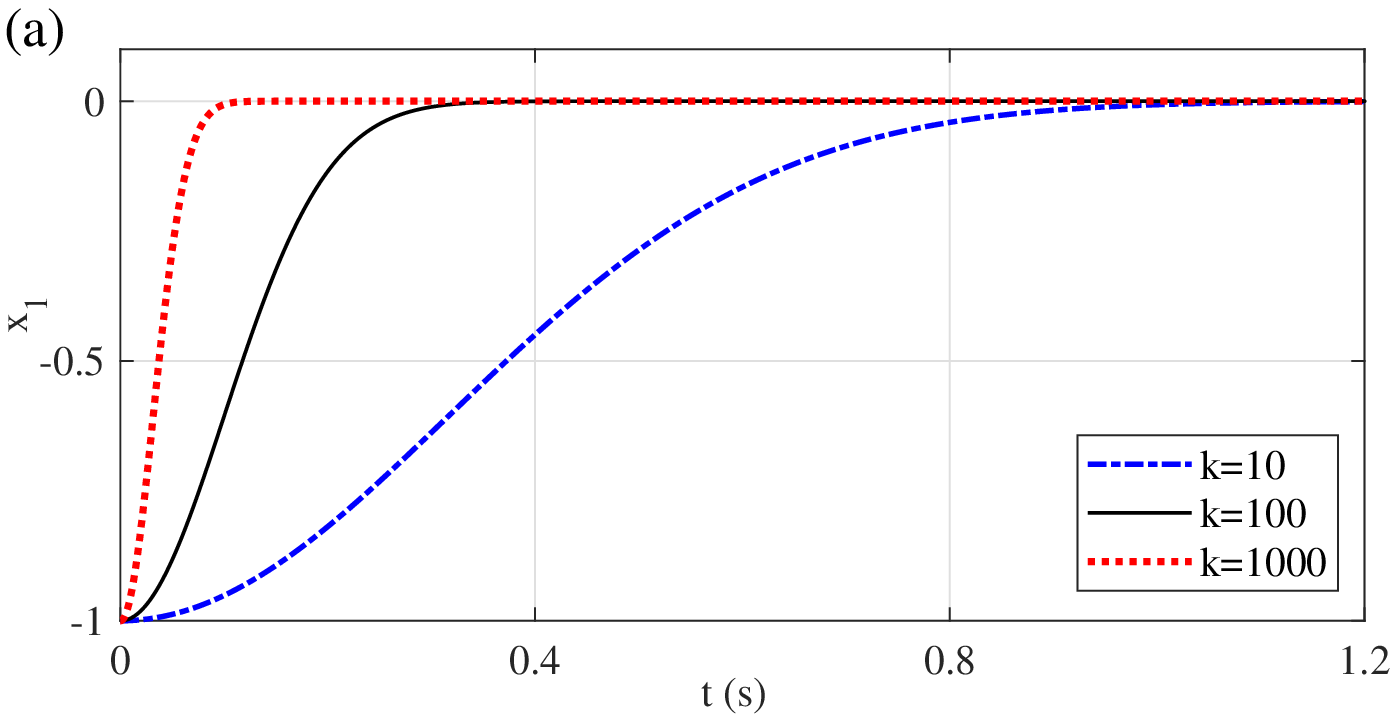}
\includegraphics[width=0.9\columnwidth]{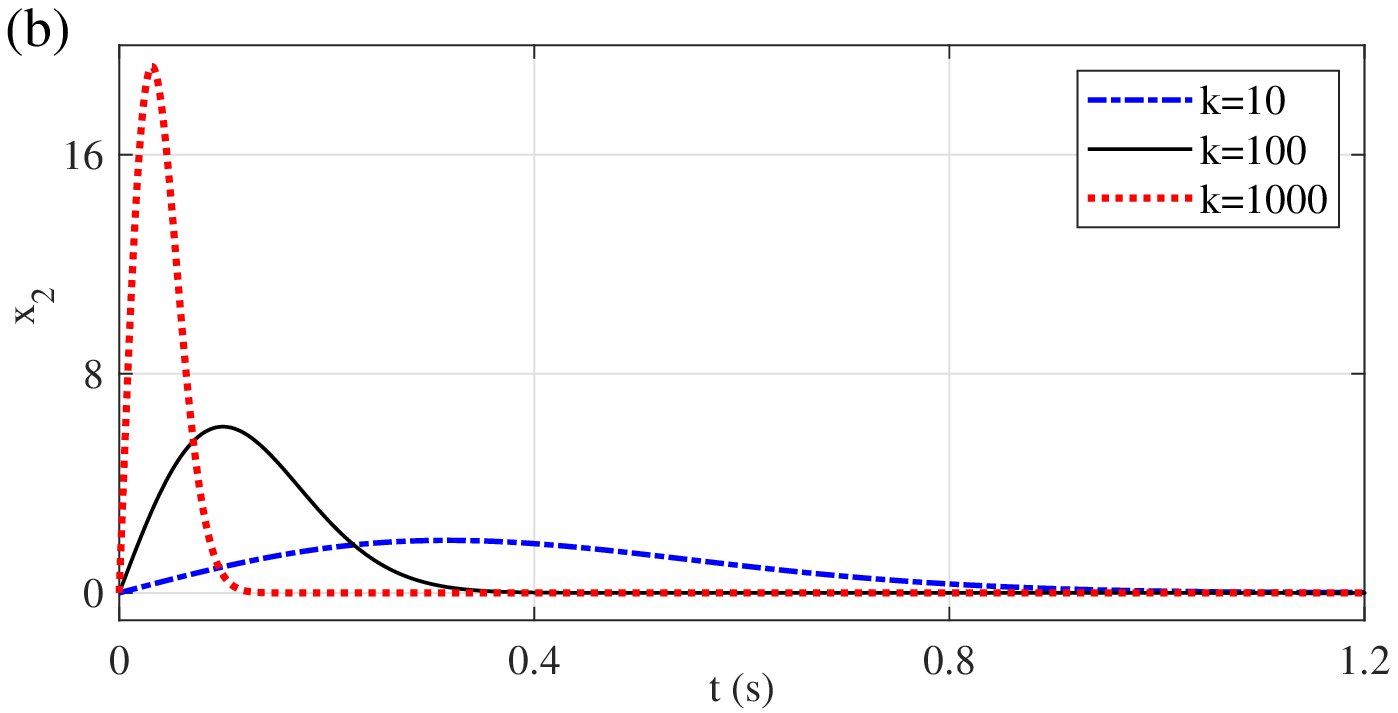}
\caption{State trajectories $x_1(t)$ in (a) and $x_2(t)$ in (b)
for the varying values of the control gain $k= [10,\,100,\,1000]$
of \eqref{eq:4}, \eqref{eq:5}.} \label{fig:2}
\end{figure}

It is also worth noting that the nonlinear damping law is
completely independent of the proportional feedback gain $k$. The
latter scales solely the state trajectories in the
$(t,x_2)$-coordinates, see examples depicted in Figure
\ref{fig:2}.

\begin{rmk}
\label{rmk:3} The closed-loop control system \eqref{eq:4},
\eqref{eq:5} allows, in addition, for a bounded control action
$|\dot{x}_2| < S$, with $S = \mathrm{const}
> 0$ to be the controller saturation (see \cite{ruderman2021}),
which is affecting yet neither the stability nor convergence
performance of the state trajectories.
\end{rmk}

Although a saturated control of \eqref{eq:4}, \eqref{eq:5} enables
bypassing singularity in the solutions of a set-point problem, the
output regulation problem requires us to allow for $x_1=0 \:
\wedge \: x_2 \neq 0$ at all times $t_0 \leq t \rightarrow
\infty$. For ensuring non-singularity of the solutions in the
entire state-space $(x_1,x_2) \in \mathbb{R}^2$, we will next make
a necessary regularization of the nonlinear damping term in
\eqref{eq:5}. With keeping this in mind, we are now in the
position to formulate the main results of this work.

\section{Main results}
\label{sec:3}

For output tracking of a reference trajectory $r(t) \in
\mathcal{C}^1$, we introduce the error state $e_1 = x_1-r$. Its
time derivative is $e_2 = x_2-\dot{r}$, respectively. Note that
for output tracking of $\mathcal{C}^1$-trajectories, one can
assume $\ddot{r}(t) = 0$ for $t > \tau$, while $t \leq \tau$
characterizes some transient phase where $\dot{r} \neq
\mathrm{const}$. In the sense of a motion control, for instance,
all $t \leq \tau$ will correspond to the transient acceleration or
deceleration phases of a reference trajectory. If a reference
trajectory $r(t)$ contains multiple, but finite in time, transient
phases with $\ddot{r}(t) \neq 0$, these will act as temporary
perturbations upon which the convergent dynamics of the control
error, i.e. $\|e_1, e_2 \| \rightarrow 0$, must be guaranteed.

With the introduced above states of the control error and the
steady-state reference (i.e. $\ddot{r}=0$), the closed-loop
control system \eqref{eq:4}, \eqref{eq:5} can be rewritten as
\begin{eqnarray}
\label{eq:7}
  \dot{e}_1 &=&e_2,  \\
  \dot{e}_2 &=&-k e_1 - \frac{|e_2| \, e_2}{|e_1| + \mu}. \label{eq:8}
\end{eqnarray}
Note that the newly introduced regularization term $0 < \mu \ll k$
does not act as an additional design parameter, yet it prevents
singularity in the solutions of the system \eqref{eq:4},
\eqref{eq:5}, cf. Section \ref{sec:2:sub:2}. Evaluating the
Jacobian of $f(x,t)$, with $x = [e_1,\,e_2]^T$ cf. \eqref{eq:7},
\eqref{eq:8} and \eqref{eq:1}, one obtains
\begin{eqnarray}
\label{eq:9}
  \frac{\partial f}{\partial x} = &
\end{eqnarray}
$$
\left[%
\begin{array}{cc}
  0 & 1 \\[2mm]
  -k + |e_2| \, e_2 \, \mathrm{sign}(e_1) / \bigl(|e_1|+\mu \bigr)^2  & - 2|e_2| / \bigl(|e_1|+\mu\bigr) \\
\end{array}%
\right].
$$
Then, suggesting the positive definite matrix
\begin{equation}\label{eq:10}
P = \frac{1}{2} \, \left[%
\begin{array}{cc}
  k & 0 \\
  0 & 1 \\
\end{array}%
\right],
\end{equation}
one can show that the matrix $J(x,t)$, which is the solution of
\eqref{eq:2}, is negative definite and, correspondingly, the
Theorem \ref{thm:1} holds. For proving it, we substitute
\eqref{eq:9} and \eqref{eq:10} into \eqref{eq:2} and evaluate the
matrix definiteness as
\begin{equation}\label{eq:11}
x^T  J(x,t) \, x = - \frac{3}{4} \, \frac{|e_2| \,e_2^2
\,\bigl(|e_1| + 2\mu \bigr)} { \bigl(e_1 + \mu \,
\mathrm{sign}(e_1) \bigr)^2} \, \leq 0 \; \; \forall \; x\neq 0.
\end{equation}
Note that the obtained inequality \eqref{eq:11} proves only the
negative \emph{semi-definiteness} of $J(x,t)$, since it is visible
that $x^T J(x,t) \, x = 0$ for $e_2 = 0  \, \wedge \, e_1 \neq 0$.
Yet it appears possible to show that $[e_1,\,e_2] = 0$ is the
unique limit solution by evaluating the $\dot{e}_2$ dynamics at
$e_2=0$. Substituting $e_2=0$ into \eqref{eq:8} results in
$\dot{e}_2 = - k e_1$. It implies that $[e_1 \neq 0, \, e_2 =
0](t)$ cannot be a limit (correspondingly steady-state) solution,
since any trajectory will be repulsed away from $e_2 = 0$ as long
as $e_1 \neq 0$. Hence, the closed-loop control system
\eqref{eq:7}, \eqref{eq:8} reveals as uniformly convergent.
Respectively, the origin of the control error $[e_1,e_2](t) = 0
\equiv \bar{x}$ is the unique limit solution, for all times $t_0 <
\tau < t$ and independent of the initial conditions $t_0$ and
$[e_1,e_2](t_0)$.

\begin{figure}[!h]
\centering
\includegraphics[width=0.48\columnwidth]{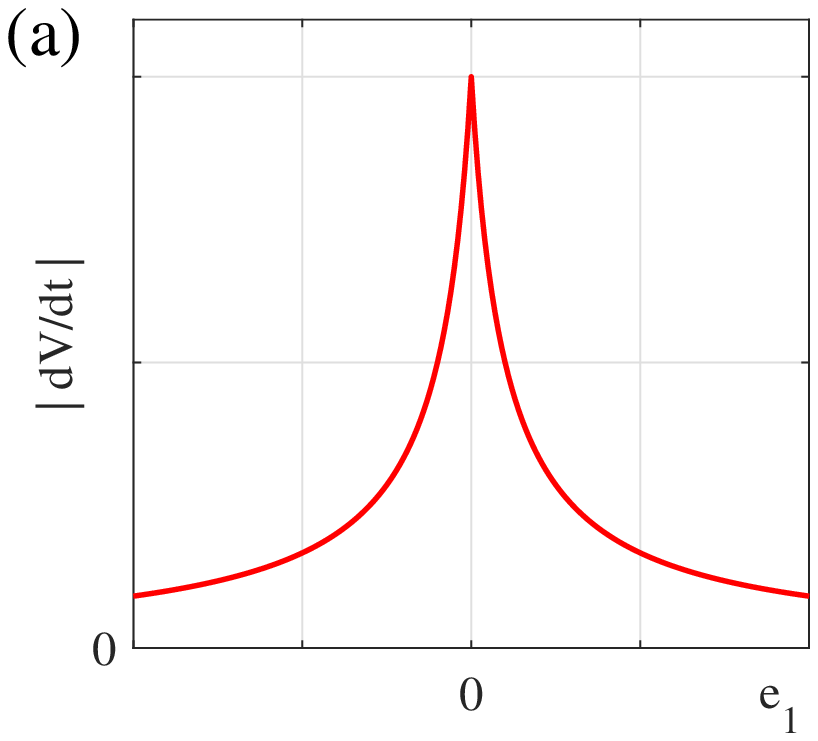}
\includegraphics[width=0.48\columnwidth]{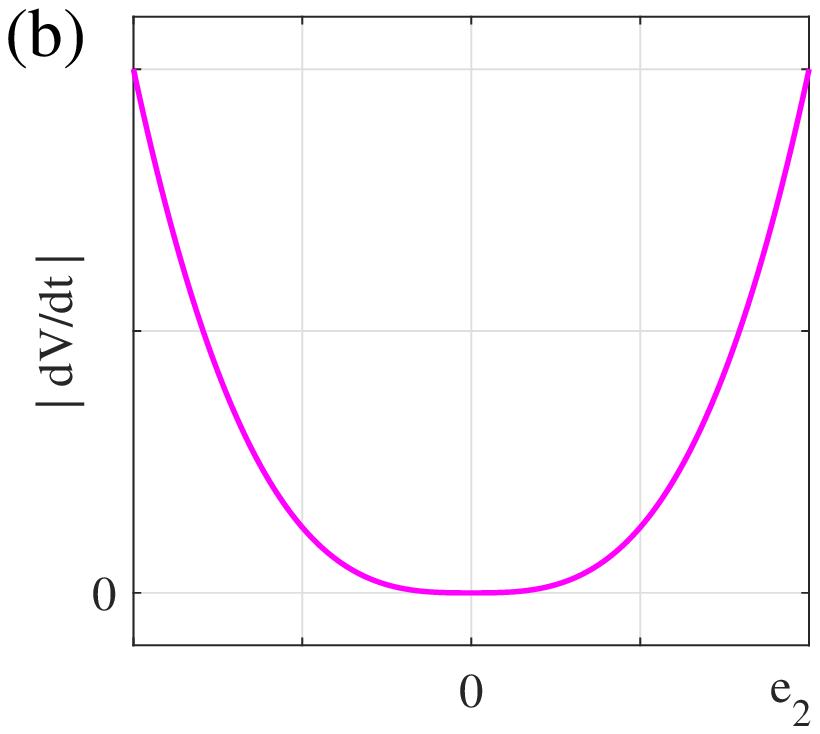}
\includegraphics[width=0.97\columnwidth]{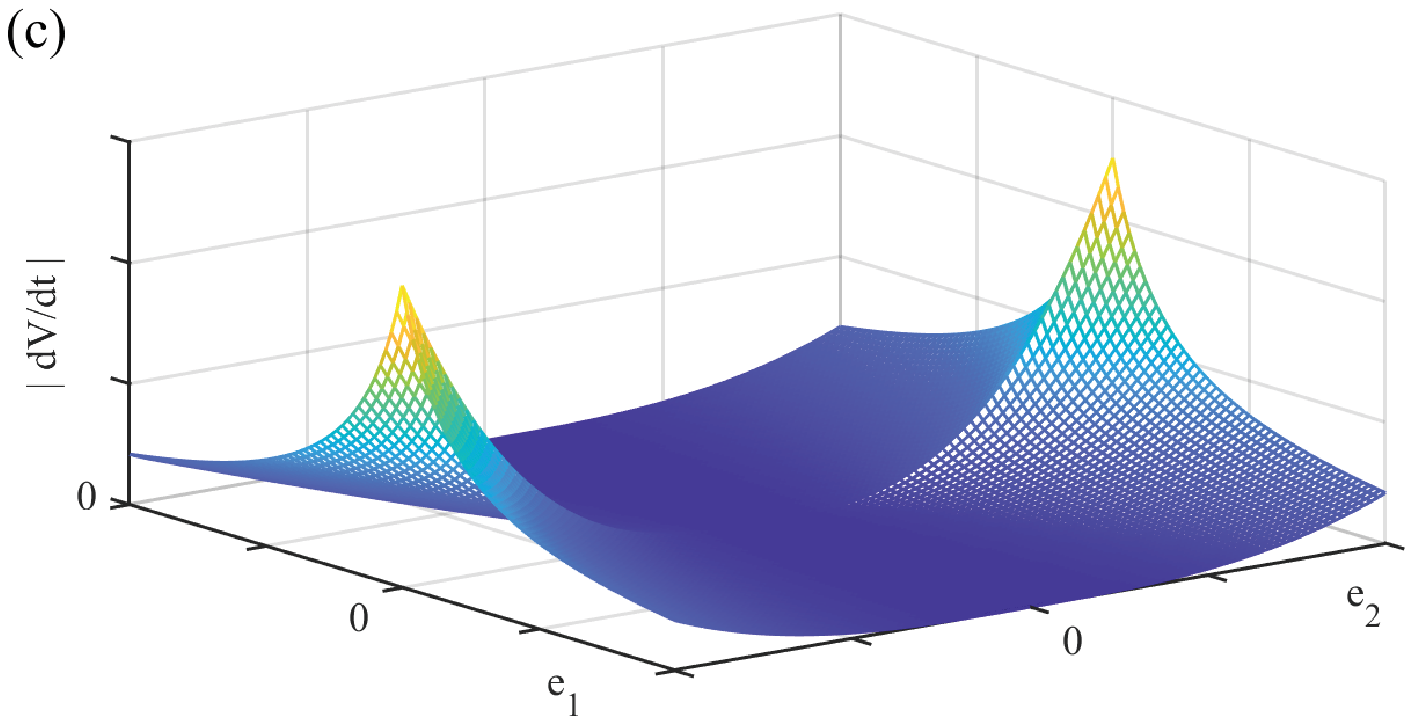}
\caption{Energy reduction rate $|\dot{V}|$ of the system
\eqref{eq:7}, \eqref{eq:8}: depending on $e_1$ in (a), depending
on $e_2$ in (b), and as the overall function, according to
\eqref{eq:13}, in (c).} \label{fig:3}
\end{figure}

\begin{rmk}
\label{rmk:4}
When assuming a quadratic Lyapunov function
candidate
\begin{equation}\label{eq:12}
V(x) = x^T  P \, x = \frac{1}{2} k e_1^2 + \frac{1}{2} e_2^2,
\end{equation}
which represents the entire energy level (i.e. potential energy
plus kinetic energy) of the system \eqref{eq:7}, \eqref{eq:8}, its
time derivative results in
\begin{equation}\label{eq:13}
\frac{d}{dt}V(x) = - \frac{|e_2|\,e_2^2}{|e_1| + \mu}.
\end{equation}
Thus, the rate at which the control system \eqref{eq:7},
\eqref{eq:8} reduces its energy is cubic in the error rate, i.e.
$\sim |e_2|^3$, and hyperbolic in the error size, i.e. $\sim
|e_1|^{-1}$, cf. Figure \ref{fig:3}.
\end{rmk}

It becomes apparent, cf. Figure \ref{fig:3} (a), that the
regularization factor $\mu$ prevents an infinite energy rate and,
thus, ensure a finite control action as $|e_1| \rightarrow 0$. At
the same time, a hyperbolic energy rate allows to accelerate the
convergence as $|e_1| \rightarrow 0$. The cubic dependence of
energy rate on the error rate enables the control to react faster
to the error dynamics. It reveals as relevant in case of, for
example, non-steady trajectory phases (i.e. $\ddot{r}(t) \neq 0$)
or sudden external perturbations which can provoke high $|e_2|$
values.

\subsection*{Numerical examples} \label{sec:3:sub:1}

Following numerical examples are provided for the implemented
system \eqref{eq:7}, \eqref{eq:8}, assuming $k=100$ and
$\mu=0.0001$ parameters and $r(t) \in \mathcal{C}^1$ reference
trajectories.

\begin{figure}[!h]
\centering
\includegraphics[width=0.49\columnwidth]{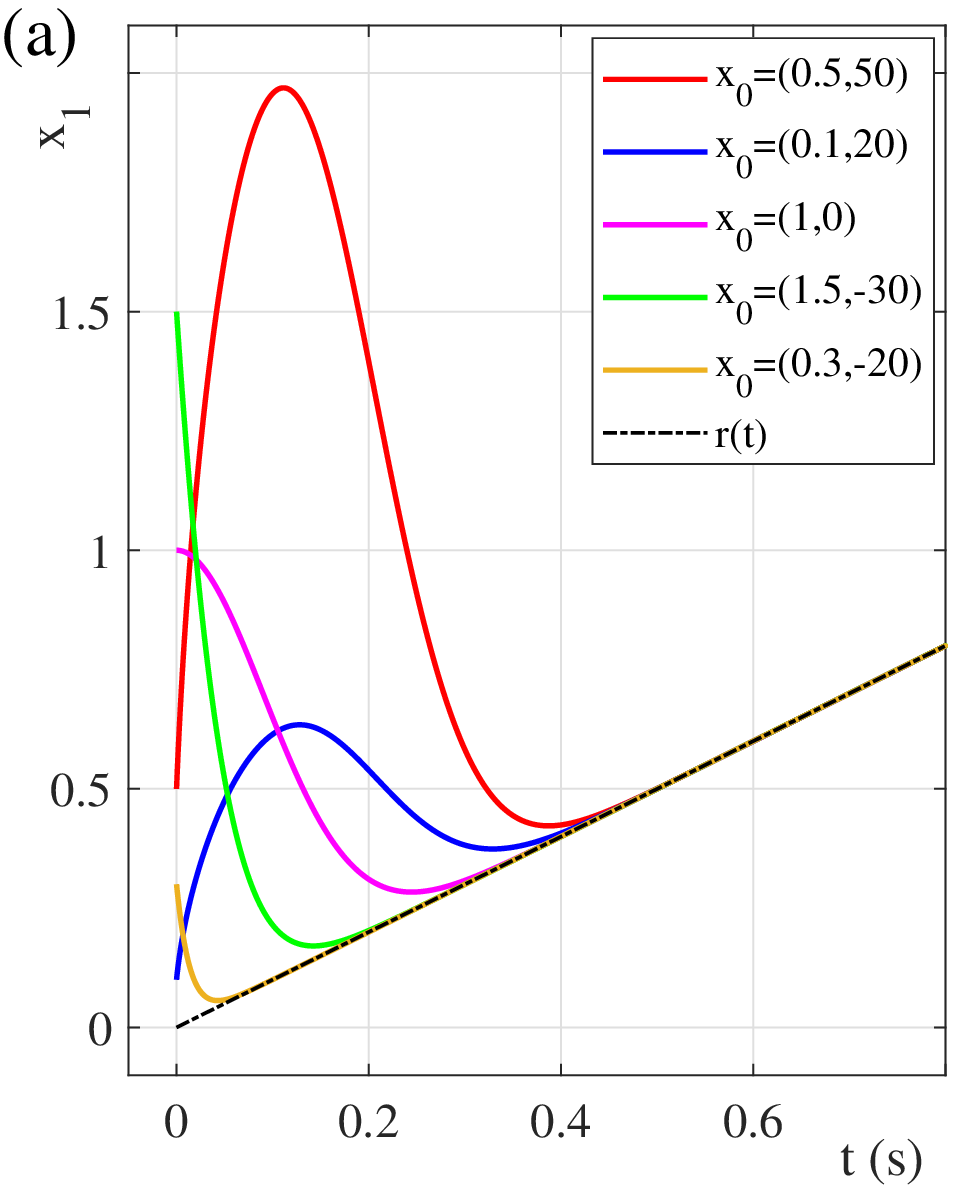}
\includegraphics[width=0.49\columnwidth]{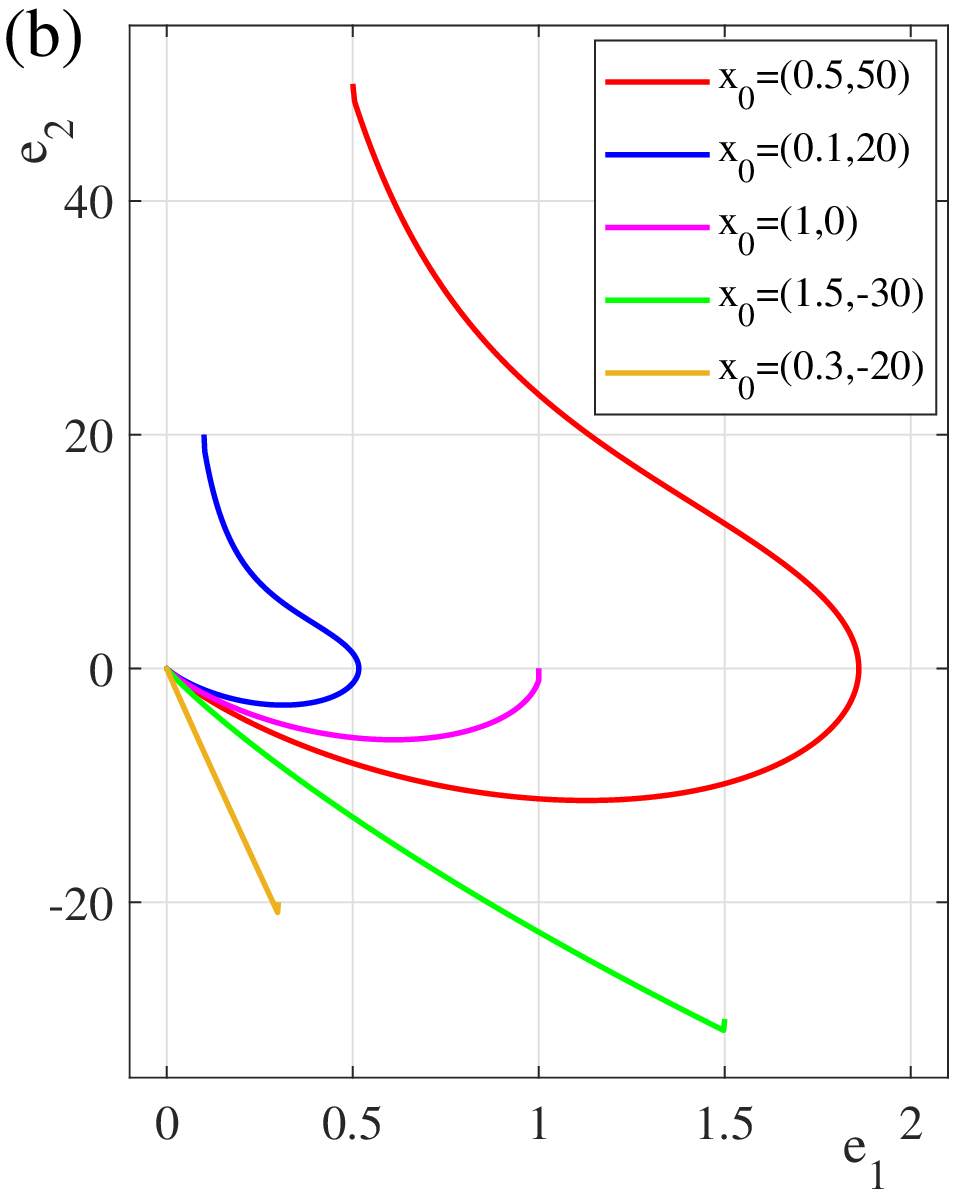}
\caption{Trajectories of the system \eqref{eq:7}, \eqref{eq:8},
with $k=100$, $\mu=0.0001$, for different initial values $x_0
\equiv [x_1,x_2](t_0)$: the output $x_1(t)$ versus reference
$r(t)$ in (a), phase portrait of the error states in (b).}
\label{fig:4}
\end{figure}

First, the output regulated trajectories are shown for the
different initial values $x_0 \equiv [x_1,x_2](t_0)$:
$$x_0 = \bigl\{ (0.5,50),\, (0.1,20),\, (1,0),\, (1.5,-30),\, (0.3,-20)
\bigr\}.
$$
The assigned reference trajectory is a linear slope $r(t) = t$.
The output response $x_1(t)$ under control is depicted in Figure
\ref{fig:4} (a). The corresponding phase portrait of the error
states, i.e. $(e_1,e_2)$, is depicted Figure \ref{fig:4} (b).

\begin{figure}[!h]
\centering
\includegraphics[width=0.98\columnwidth]{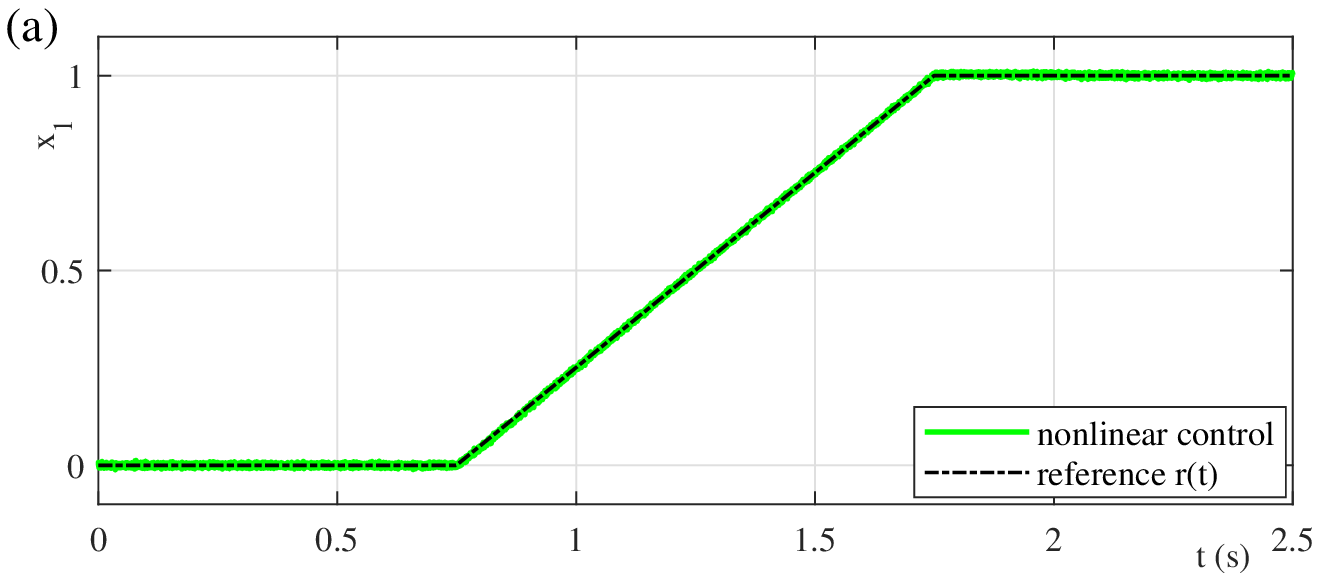}
\includegraphics[width=0.98\columnwidth]{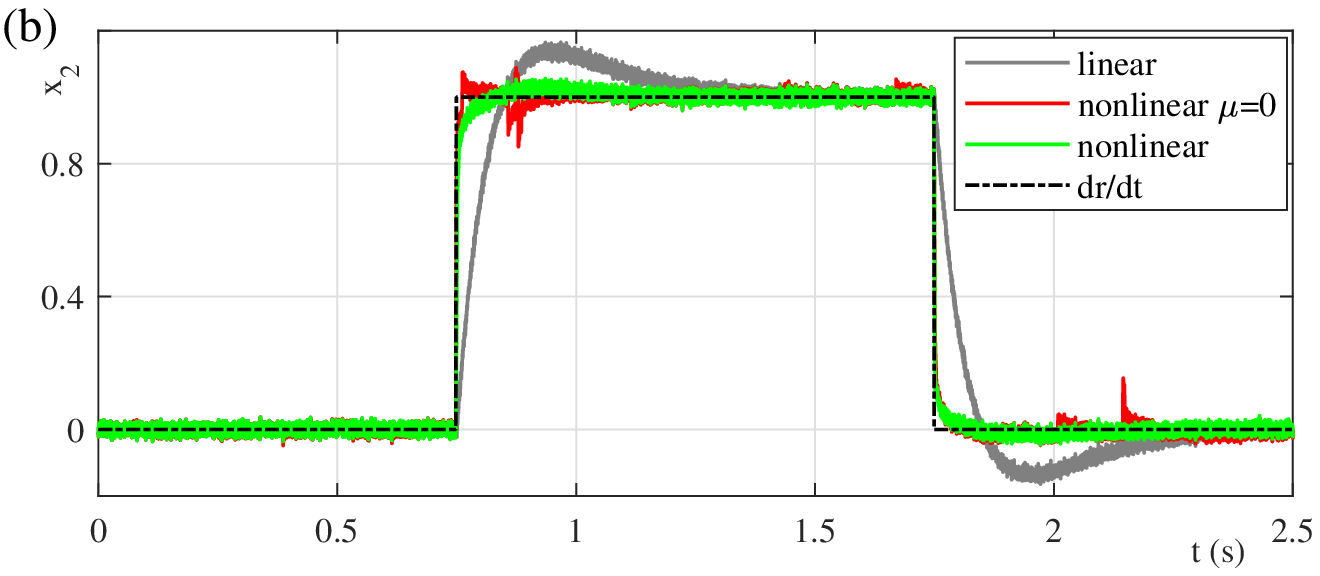}
\caption{Trajectories of the system \eqref{eq:7}, \eqref{eq:8},
with $k=100$, $\mu=0.0001$: the output $x_1(t)$ versus reference
$r(t)$ in (a), the $x_2(t)$ state in (b) -- compared with case
without regularization (i.e. $\mu=0$) and also with critically
damped proportional-derivative linear control.} \label{fig:5}
\end{figure}

Next, we demonstrate the control performance of the output
tracking when $r(t)$ is only piecewise $\mathcal{C}^1$ and
contains the finite phases where $\dot{r}(t) \neq \mathrm{const}$.
Furthermore, in order to emphasize a practical control
applicability, both $x_1(t)$ and $x_2(t)$ signals, used for the
feedback control in \eqref{eq:8}, are subject to a non-correlated
bandlimited white-noise. The output response $x_1(t)$ under
control is depicted in Figure \ref{fig:5} (a) over the reference
trajectory.
The $x_2(t)$ state, which corresponds to the relative velocity if
(for example) a motion control is intended, is depicted in Figure
\ref{fig:5} (b). For the sake of comparison, the case without
regularization, i.e. $\mu=0$, is also shown. Moreover, the
$x_2(t)$-response of a critically damped proportional-derivative
linear control is also demonstrated, for the sake of comparison.
The assigned linear controller, with the same $k=100$, features
the $\dot{e}_2 = -100 e_1 - 20 e_2$ error dynamics,
correspondingly.

\section{Conclusions}
\label{sec:5}

The Demidovich's concept of convergent system dynamics was used
for analyzing the convergence properties of the optimal nonlinear
damping control, introduced in \cite{ruderman2021}. The existence
of unique limit solution in the coordinate origin of the output
tracking error and its time derivative was demonstrated. A
regularization term was introduced, comparing to
\cite{ruderman2021}, which prevents singularity in the solutions
when an output zero crossing occurs outside of origin. The
provided analysis and results can be relevant for using the
optimal nonlinear damping control as an alternative to a standard
proportional-derivative controller. The optimal nonlinear damping
control performs as significantly faster converging and is,
moreover, robust against the measurement noise of the feedback
states.

\bibliographystyle{elsarticle-num}        

\bibliography{references}

\end{document}